# Ferroelectric Properties of Monoclinic Pb(Mg$_{1/3}$Nb$_{2/3}$)O$_3$ – PbTiO$_3$ Crystals

## A. A. Bokov and Z.-G. Ye[*]

*Department of Chemistry, Simon Fraser University, Burnaby, BC, V5A 1S6, Canada*

A monoclinic phase was recently discovered near the morphotropic phase boundary in several high-performance piezoelectric perovskite solid solutions, but its properties have not been reported. In this paper the dielectric, piezo- and ferroelectric properties of the monoclinic $Pm$ phase in the $(1-x)$Pb(Mg$_{1/3}$Nb$_{2/3}$)O$_3 - x$PbTiO$_3$ perovskite system are studied. In a (001)-oriented crystal of composition $x \approx 0.33$, ferroelectric hysteresis loops with remanent polarization of 23 $\mu$C/cm$^2$ are displayed. In poled monoclinic crystals, under unipolar drive up to 10 kV/cm, the domain walls remain unchanged, the polarization and longitudinal strain change almost linearly, but the piezoelectric response ($d_{33}$=9×10$^{-10}$ C/N) is much weaker than in the rhombohedral phase of close composition. The relative dielectric permittivity of the $Pm$ phase is also smaller (with a small-signal value of ~ 2500), but the piezoelectric constant ($g_{33} = 3 \times 10^{-2}$ m$^2$/C) and the electromechanical coupling factor ($k_t = 0.60$) are practically the same as in the rhombohedral phase. The properties of the various phases in the range of the morphotropic phase boundary are related to the different rotation paths of the polarization vector induced by the external drive.

PACS numbers: 77.80.Dj, 77.65.-j, 77.22.-d , 77.84.Dy

## I. INTRODUCTION

Ferroelectric solid solutions with compositions close to the morphotropic phase boundary (MPB) are widely used in advanced technology because of their extraordinary properties. MPB is defined as a compositional dividing line between two adjacent phases in a temperature ($T$) vs composition ($x$) phase diagram. The best known example is the perovskite ferroelectric system $(1-x)$PbZrO$_3 - x$PbTiO$_3$ (PZT) in which an almost temperature-independent MPB is observed at $x \cong 0.5$. For many years pure and modified PZT ceramics had been the main materials for piezoelectric devices and the subject of intensive experimental and theoretical investigations.[1] It was initially believed that the MPB in the PZT system separates a rhombohedral and a tetragonal phases. Recently, structural studies by means of synchrotron x-ray powder diffraction technique revealed by surprise a monoclinic $Cm$ phase with a narrow composition range lying in between the rhombohedral $R3m$ and the tetragonal $4mm$ phases.[2]

Other exciting results have more recently been obtained in the perovskite crystals of the $(1-x)$Pb(Zn$_{1/3}$Nb$_{2/3}$)O$_3$ - $x$PbTiO$_3$ (PZNT) and $(1-x)$Pb(Mg$_{1/3}$Nb$_{2/3}$)O$_3$ - $x$PbTiO$_3$ (PMNT) systems. Similar to PZT, a rhombohedral $R3m$ and a tetragonal $P4mm$ phases exist in these solid solutions at low and high $x$, respectively, and the compositions close to the MPB exhibit enhanced electromechanical properties, with a very high piezoelectric constant $d_{33}$ > 2000 pC/N, and a very large electromechanical coupling factor $k_{33}$ > 90%. The piezoelectric properties of the PZNT and PMNT crystals outperform the PZT ceramics, leading to a revolution in electromechanical transducer technology.[3,4,5] X-ray and neutron diffraction studies have also revealed in these systems an intermediate monoclinic phase between the known $R3m$ and $P4mm$ regions.[6,7,8,9] In PMNT, the new phase was found at $x \cong 0.35$ and the width of its existence range $\Delta x$ was estimated to be about 0.03. The monoclinic phase was also observed using polarized light microscopy.[10] Theoretically, it was derived from the phenomenological Devonshire approach.[11]

Since the intermediate monoclinic phase has been found in the range of MPB in several high-performance piezoelectric materials, a natural presumption is that, it is the monoclinic phase that gives rise to the extraordinary piezoelectric properties.[12] However, no direct measurements of the ferroelectric and electromechanical properties of that new phase have been reported so far. The purpose of this paper is to present the original results on the electrical behavior of the monoclinic phase.

One of the common difficulties encountered in the study of perovskite solid solution crystals is the unavoidable macroscopic spatial variations of the cation ratio on the B-site, which occurs during the crystal growth. For instance, an elementary analysis by laser ablation inductively coupled plasma mass spectrometry showed that in PMNT crystal, the $Ti^{4+}$ local concentration, $x$, may vary up to ±5% from its nominal composition,[13] which is comparable with the estimated composition range of the monoclinic phase. For the crystals with composition close to the MPB, the gradients of $x$ may lead to the coexistence of different ferroelectric phases in the same specimen, which was actually observed by several authors (see e.g. Refs. 14, 15). The mixture of phases was also reported for the ceramic PMNT samples with the compositions near the MPB.[9,16] Therefore, any meaningful studies of relationship between the structure and properties of PMNT and similar solid solutions should be complemented by careful examinations of the crystal symmetry and phase components of the very same sample in which the measurements of properties are to be performed. Note that X-ray diffraction and related techniques, which are commonly used in practice, are not fully suitable for these purposes, because the lattice distortions in different phases differ only slightly from each other, and a small admixture of the secondary phase may not be detected, but may affect greatly the physical properties. That is why the measurements in this work are accompanied by simultaneous observation of the domain structures by polarization light microscopy. In this way, the tetragonal, rhombohedral and monoclinic phases, that are expected to exist for the compositions near the MPB, are unambiguously identified and distinguished from each other.

---

[*] Corresponding author, Tel: (604) 291-3351, Fax: (604) 291-3765, email: zye@sfu.ca

## II. EXPERIMENTS

Single crystals of $(1-x)Pb(Mg_{1/3}Nb_{2/3})O_3 - xPbTiO_3$ with nominal composition $x = 0.35$ were grown by the Bridgman method. Since the highest electromechanical coupling is known to occur in PMNT (as well as in PZNT) crystals when poled along the <001> pseudocubic directions,[3-5] (001)-oriented plate-like specimens were prepared with the help of Laue camera (all indexes are referred to the cubic system). The large faces of the crystal plate were mirror polished with gold electrodes sputtered on for the electrical characterization. The poling was performed by an electric field of 10 kV/cm applied at room temperature. The variations of polarization and strain versus electric field were measured using a Radiant RT66A Test System and a fiber-optic system MTI-2000, respectively. A drive voltage of triangular pulses was applied. The sample holder was designed to allow the crystal to deform without mechanical constraints. The dielectric permittivity in the range of $10^{-2}$ – $10^5$ Hz and electromechanical resonance frequencies were determined using a Solartron 1260 impedance analyzer and a Solartron 1296 dielectric interface. For *in situ* study of the domain structure under a dc bias, semitransparent gold layers were sputtered as electrodes. Gold wires were attached to the electrodes by silver paste to connect the sample with a high voltage source. The domain structures were studied by polarized light microscopy. The direction of the polarized light propagation and that of the applied electric field were parallel to each other, and to the [001] direction of the crystal.

## III. RESULTS AND DISCUSSION

Preliminary examinations of large (~10 x 10 $mm^2$) crystal plates in polarizing microscope revealed the coexistence of macroscopic domains of different phases, namely, the rhombohedral and tetragonal phases, and another phase of lower symmetry. We successfully identified the part of the crystal containing the low-symmetry phase only, cut a few (001)–oriented monophase plates with an area of ~2 × 2 $mm^2$ and a thickness of 0.14 mm, and studied their properties. All regions of the platelets showed in crossed polarizers a clear extinction that is not parallel to <100> or <110>. Such an optical behavior is not compatible with either a tetragonal or a rhombohedral symmetry. More detailed analysis of the domain structure of these samples before and after poling (which will be discussed in a separate paper) revealed a monoclinic *Pm* symmetry, in agreement with the neutron [8] and x-ray [9] diffraction data (the phase of this symmetry is also called $M_C$ phase). Upon heating, a change in the domain structure was observed at 82 - 88 °C (with the different parts of crystal showing different temperatures), indicating a phase transition into another (ferroelectric) phase, which, according to the published PMNT phase diagram,[7,16,21] should be the tetragonal *P4mm* phase. At 150 - 155 °C, a second transition into the high-symmetry (cubic) phase occurs and the crystal becomes optically isotropic. The variation of phase transition temperature across the crystal can be explained, as discussed above, by the spatial variation of the composition $x$. The temperature dependence of the real part of dielectric permittivity $\varepsilon'$ showed a maximum at $T_m = 162$ °C. This temperature was used to estimate the average composition $x_{av}$ of the sample by comparing it with the $T_m$ values of the ceramics, where the composition is known exactly and $T_m(x)$ is a linear function.[16,17,18] In this way, it was found that $x_{av} \approx 0.33$.

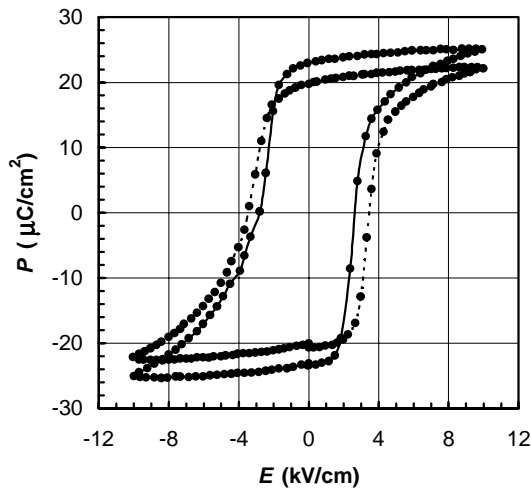

FIG. 1. Hysteresis loops displayed in a monoclinic PMNT (001)-oriented crystal at 25 ºC at frequencies of 1 Hz (solid line) and 100 Hz (dashed line).

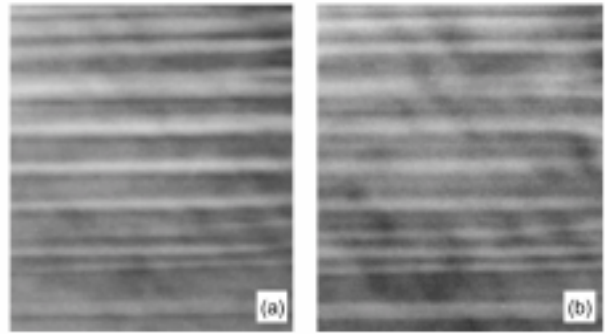

FIG. 2. Domain structure of the monoclinic phase observed on the (001) PMNT platelet: (a) under an electric field of 10 kV/cm // [001] (i.e. perpendicular to the plane of the platelet); (b) after the removal of the field.

The dielectric hysteresis loops at a drive voltage of different frequencies are shown in Fig. 1. The well-saturated and symmetrical loops indicate the ferroelectricity of the monoclinic phase. The remanent polarization $P_r$ and the coercive field $E_c$ depend on frequency when it is higher than about 1 Hz, but in the low-frequency range the variation becomes negligible, e.g., the same values of $P_r = 23$ $\mu C/cm^2$ and $E_c = 2.7$ kV/cm are obtained at 1 Hz and 0.2 Hz. This means that the process of domain switching are practically completed during the period. As the spontaneous polarization vector of the $Pm$ phase lies somewhere in between [001] and [101], the magnitude of its spontaneous polarization $P_s$ is estimated to be: $P_r < P_s < \sqrt{2}\, P_r = 33$ $\mu C/cm^2$.

Fig. 2 gives the photograph of the domain structure of the crystal observed by polarizing microscope. It consists of laminar birefringent domains separated by straight dark boundaries, which are oriented along <110>. The width of the domain stripes is about 1-4 μm. As the spontaneous polarization vectors of all the domains in the poled monoclinic phase form the same angle to the [001] direction, the change of energy density caused by the electric field applied afterwards in that direction should be the same for all the domains. Consequently, such a field should not affect the domain walls. In agreement with this analysis, no any noticeable changes in the configurations of the domain walls are observed under the electric field [compare Figs. 2 (a) and (b) as an example].

The dielectric spectra measured at room temperature under a small (3 V/cm) ac signal are shown in Fig. 3. A significant dispersion is evidenced in the whole measurement range of frequency, suggesting an extremely wide distribution of relaxation times. After poling, the dispersion is attenuated and the real part of permittivity diminishes dramatically. The stronger dielectric dispersion and the higher values of permittivity and tan$\delta$ in the unpoled state can be attributed to the motion of the domain walls, which is a usual phenomenon in multi-domain ferroelectrics. As observed above, a field does not affect the position of the domain walls of the poled crystal. As a result, they no longer contribute to the dielectric response. Another possible cause for the change of the dielectric properties after poling is the anisotropy of monoclinic phase. The losses at very low frequencies still remain significant after poling. They are probably arise from the polarization of mobile charge carriers, characteristic of many materials at low frequencies including PMNT.[19]

The electromechanical coupling of the poled crystal is measured by the IEEE resonance technique.[20] The value of $k_t$ is found to be 0.60, which is within the range of the values (0.54 – 0.62) reported for the PMNT crystals of MPB composition.[3,15]

The dependences of the polarization and the longitudinal strain $S$ on the unipolar drive field in the poled crystal are shown in Fig. 4. Both of them are almost linear and nearly nonhysteretic. From these dependences, the piezoelectric constants, $d_{33} = S/E \approx 900$ pC/N and $g_{33} = S/P \approx 30\times 10^{-3}$ $m^2$/C, are found. This value of $d_{33}$ falls into the interval of 340-2800 pC/N previously reported for different (001)- oriented

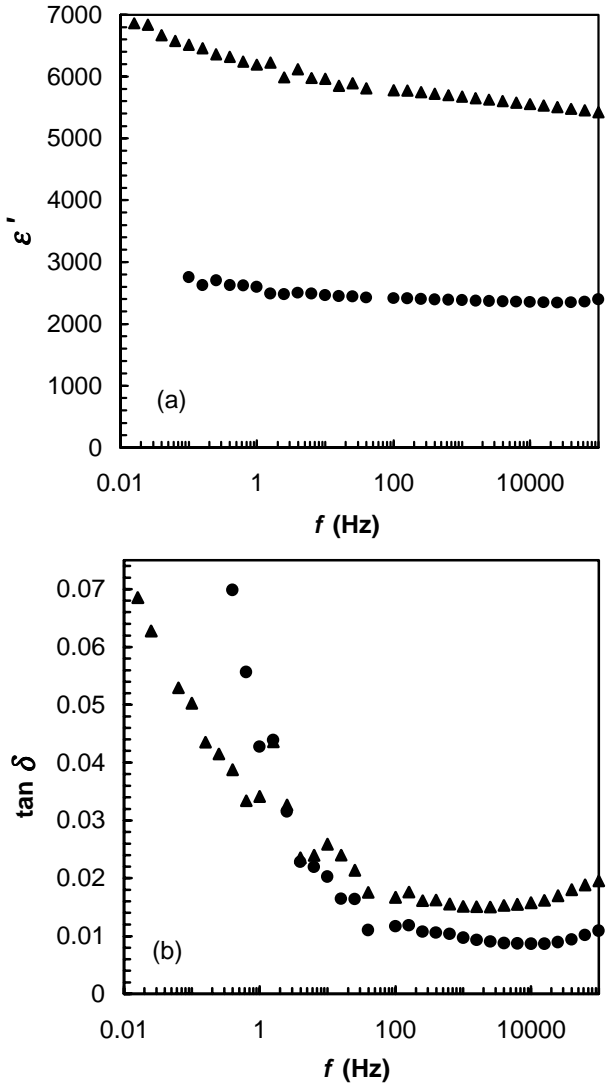

FIG. 3. Frequency dependences of (a) the relative permittivity, and (b) tan$\delta$ of the (001)-oriented monoclinic PMNT crystal measured at 25 $^{\circ}$C, before (triangles) and after (circles) poling.

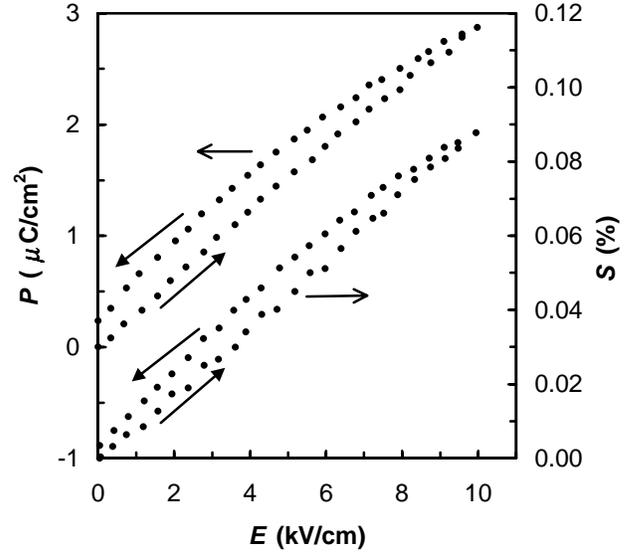

FIG. 4. Polarization (upper curve) and longitudinal strain (lower curve) as a function of unipolar drive field ($f = 1$ Hz) measured in the (001)-oriented and poled monoclinic PMNT crystal.

PMNT crystals with the MPB composition [3,15,21] and for different regions of the same crystal,[22] but much smaller than the upper limit value of that interval. The same can be said regarding the relative permittivity. We find in the poled crystal the small-signal value of $\varepsilon'_{33} \approx 2500$ [see Fig. 3 (a)], while the reported values for different MPB compositions are about 3500-5500.[3,15] The significantly different values of $d_{33}$ (and $\varepsilon'_{33}$) measured in the crystals with the same or close nominal composition can be attributed to the presence of different phases near the MPB, i.e. the rhombohedral phase that has the highest piezoelectric constant, the monoclinic phase with a smaller constant, and probably the tetragonal phase. The authors of Ref. 21 reported $d_{33}$ values in the range of $900 - 1100$ pC/N for the PMN-xPT crystals with $x = 0.35$, which agree well with our results. It seems that their crystal might inadvertently be composed (or mainly composed) of the monoclinic phase. In the other study of the PMNT crystals with MPB composition,[22] the magnitude of the piezoelectric constant was related to the domain configuration. It was shown that the crystals with large ($> 0.1$ mm) domains had a higher $d_{33}$ value ($1700 - 2800$ pC/N), while small laminar domains (resembling those observed in our work) had a smaller $d_{33}$ value ($340 - 1810$ pC/N). It is possible that the large and small domain regions belong to the distinct rhombohedral and monoclinic phase, respectively, which would explain the difference in the $d_{33}$ values measured.

On the other hand, in the PMNT crystals of MPB having large $d_{33}$ (i.e. in the rhombohedral phase), the value of $g_{33} = 2.73\times 10^{-2}$ m$^2$/C was reported,[23] which is practically the same as what we have found in the monoclinic phase. In the case of the experimental set-up used in our work, one can write $d_{33} \approx \varepsilon_0\varepsilon'_{33}g_{33}$. The equality of $g_{33}$ in both the monoclinic and the rhombohedral phases means that the superior $d_{33}$ constant of the latter is not due to the enhanced coupling between $P$ and $S$ (which is characterized by $g_{33}$), but due to the large dielectric response (i.e. high $\varepsilon'_{33}$). This is consistent with the explanation of the exceptional piezoelectric properties of the rhombohedral $R3m$ phase, by an easy electric field induced rotation of the polarization vector within the (110) plane from [111] (the direction of the polarization in the rhombohedral phase at zero field) to [001] (the direction of the applied field).[4] According to the first-principle calculations performed for the perovskite structure,[24] it is this polarization rotation path (designated as $a \rightarrow f \rightarrow g$ in Fig. 2 of Ref. 24) that provides the flattest energy surface, so that an electric field along [001] causes a large change of polarization angle, a large increase in the polarization component along [001] and thereby a large piezoelectric response along that direction. The other consequence of such kind of energy profile is that, a [001]-field should change the rhombohedral symmetry into the monoclinic $Cm$ (or in other name, $M_A$) symmetry, in which the polarization remains in the (110) plain. Thus, under this field, the polarization vector in rhombohedral and $Cm$ phases follows the same path and a large $d_{33}$ value of the same magnitude can be expected for the both cases. In the monoclinic $Pm$ phase, however, under a [001]-field, the polarization vector has to rotate within another plane, i.e. (010) ($c \rightarrow d$ path in Fig. 2 of Ref. 24). At zero temperature where the calculations were performed,[24] this process requires much more energy, and thus is less favorable for the piezoelectric response. The corresponding first-principle calculations for finite temperatures have not been published, but it is sensible to expect that the relative energy profiles experienced by different polarization paths remain qualitatively the same. Thus, $d_{33}$ in the rhombohedral crystal appears to be much larger than in the monoclinic $Pm$ crystal with a slightly larger concentration of $x$.

The same arguments can be applied to the other analogous systems, in particular to the PZN-PT solid solution in which the same phase sequence of rhombohedral, intermediate monoclinic $Pm$ and tetragonal structures was observed at room temperature with increasing $x$ [8] (the orthorhombic $Bmm2$ symmetry of the intermediate phase was reported in Ref. 6, but it can also be regarded as a particular case of monoclinic $Pm$ phase with the lattice parameters $a=c$). The published $d_{33}$ ($x$) dependence of this system [3,4] shows a maximum value of $\sim$ 2500 pC/N at $x$ = 7 – 8%, i.e. in the rhombohedral side of the phase diagram. In the MPB region, $d_{33}$ decreases abruptly (to 880 – 1600 pC/N at $x$ = 9.5 %), which can be explained by the presence of the monoclinic (or orthorhombic) phase. Further decrease of $d_{33}$ (down to $\sim$ 500 pC/N) observed at higher $x$ is clearly due the transition into the tetragonal phase, in which the direction of polarization coincides with that of the applied electric field, and thus the polarization rotation process no longer takes place. Similar to the case of PMNT, the permittivity $\varepsilon'_{33}$ of the rhombohedral phase in PZNT ($\sim$ 4500 at $x$ = 8 %) is larger than that of the of the monoclinic phase ($\sim$ 1500 at $x$ = 9.5 %) and the value of the coupling factor does not depend significantly upon the phase content ($k_t$ equals 0.48 and 0.54 at $x$ = 8 and 9.5%, respectively).[3]

On the other hand, the properties of the (111)-oriented PZNT crystals show a reversed composition-dependence:[3,4] the values of the piezoelectric coefficient ($d_{33}$ = 82 pC/N) and the permittivity ($\varepsilon'_{33}$ = 2150) of the rhombohedral composition ($x$ = 8%) are much smaller than those ($\sim$ 500 pC/N and 4300, respectively) of the monoclinic composition ($x$= 9.5%). This behavior can also be understood in terms of the above-mentioned polarization rotation mechanisms. In the rhombohedral phase, the electric field is applied in the same direction as that of the spontaneous polarization $\mathbf{P}_s$ (i.e. along <111>) and therefore, it cannot rotate $\mathbf{P}_s$. In the monoclinic phase, the polarization rotation process takes place, giving rise to the enhanced permittivity (coupling between $P$ and $E$) and $d_{33}$ (strain per unit field). But the polarization rotation path is different from that in the (001) crystals; it is neither within (110) nor within (010) plane. As a result, the resulting piezoelectric response in the (111) crystals is not as strong as in the (001) crystals.

The piezoelectric effect in PZT was studied for ceramics only, because of the lack of good quality single crystals. But the properties of this system in the MPB range were predicted theoretically using an *ab initio* approach.[25] Application of an electric field along [001] was predicted to lead to the transformation of the rhombohedral phase into the monoclinic $Cm$ and then the monoclinic $Pm$ phase. Once again, the $d_{33}$ value in the rhombohedral and $Cm$ phases, associated with a polarization rotation path within the (110)-plane, was calculated to be much higher than that in the $Pm$ phase with a rotational path within the (010)-plane.

## IV. CONCLUSIONS

We have characterized the dielectric, piezoelectric and ferroelectric properties of the monoclinic $Pm$ phase of the PMNT crystals, and found that the piezoelectric response is not as strong as that observed in the crystals containing the rhombohedral phase with a composition near the MPB. Although the measurements were made along the [001] direction only, one can hardly expect that the results would be better in the other directions. The peculiarities of the piezoelectric properties of the monoclinic phase can be attributed to its particular symmetry and the related polarization rotation pathway, which seems to be less energetically favorable than the rotation of the polarization in the rhombohedral phase. Our results and analysis lead to the following observation. If an electric field applied to a poled crystal along the [001]-direction causes the rotation of its polarization vector within the {110} planes (such as in the rhombohedral or monoclinic *Cm* phase), the piezoelectric response appears to be larger than in the case where the polarization rotation takes place in {010} planes (e.g. in orthorhombic or monoclinic *Pm* phase). If the [001]-field cannot rotate the polarization at all (i.e. in the case of the tetragonal phase), the piezoelectric response will be the minimum. Such a dependence of the piezoelectric properties upon the paths of field-induced polarization rotation seems to be characteristic to all the MPB-related ferroelectric solid solution systems of perovskite structure.

More extensive studies of the nature of the monoclinic phase are highly desirable, both in theoretical and experimental aspects. On the other hand, for the technological applications, it is necessary to avoid using the crystals of the monoclinic *Pm* phase and to select the rhombohedral side of the MPB in order to ensure the highest piezoelectric constant $d_{33}$. The optical control of the domain states in the piezoelectric crystals appears to be very important in such studies and applications.

## ACKNOWLEDGMENTS

The authors very grateful to Dr. H. Luo for help in crystal preparation. This work was supported by the U.S. Office of Naval Research (Grant No. N00014-99-1-0738).